\documentclass[twocolumn,showpacs,preprintnumbers,amsmath,amssymb]{revtex4}

\usepackage{graphicx}
\usepackage{dcolumn}
\usepackage{bm}
\usepackage{rotating}

\begin{document}

\preprint{APS/123-QED}

\title{Fast magnetic reconnection \\in three dimensional MHD simulations}

\author{Bijia Pang}
 \email{bpang@physics.utoronto.ca}
 \affiliation{Department of Physics, University of Toronto, Toronto ON, M5S 1A7, Canada}

\author{Ue-Li Pen}%
 \email{pen@cita.utoronto.ca}
 \affiliation{Canadian Institute for Theoretical Astrophysics, University of Toronto, Toronto ON, M5S 3H8, Canada}

\author{Ethan T. Vishniac}
 \email{ethan@mcmaster.ca}
 \affiliation{Department of Physics and Astronomy, McMaster University, Hamilton ON, L8S 4M1, Canada}%

\date{\today}

\begin{abstract}

We present a constructive numerical example of fast magnetic reconnection in a three
dimensional periodic box.  Reconnection is initiated by a strong, localized perturbation to the field lines.
The solution is intrinsically three
dimensional, and  its gross properties do not depend on the
details of the simulations.  $\sim 50\%$ of the magnetic energy is released
in an event which lasts about one Alfven time, but only after a delay during which the field lines evolve into a critical configuration.  We present a physical picture of the process.  The
reconnection regions are dynamical and mutually interacting.   In the comoving frame of these
regions, reconnection occurs through an X-like point, analogous
to Petschek reconnection.  The dynamics appear to be driven by global
flows, not local processes.

\end{abstract}

\pacs{52.30.Cv}
\keywords{magnetic reconnection}
\maketitle

\section{Introduction}

Most of the matter in the universe exists in the plasma state and plasma also
plays an important role in  gas dynamics in astrophysics.  When
magnetic field are present, the dynamics of an electrically conducting plasma
is sensitive to magnetic  forces; as a result,
magnetohydrodynamics(MHD) is used to understand the dynamical evolution of astrophysical
fluids.

The ideal limit of MHD poses a new class of problems in dissipative processes.  In ideal
hydrodynamics, irreversible processes, such as shock waves and vorticity
reconnection, occur at dynamical speeds, independent of microscopic
viscosity parameters.  Weak solutions describe these irreversible
discontinuous solutions of the Euler equations.  While smooth flows
conserve entropy and vorticity, the infinitesimal discontinuity surfaces
generate entropy and reconnect vorticity.  This can also be understood
as a limiting case starting with finite viscosity, where these surfaces
have a finite width.

With magnetic fields, a more dramatic problem emerges.  If two
opposing field lines sit nearby, a state of higher entropy  can be reached by reconnecting
the field lines, and converting their magnetic energy into fluid entropy.  
In the presence of resistivity, this process occurs on a resistive
time scale for some relevant scale. This exaggerates the problem somewhat.  
Extensive theoretical research on magnetic
reconnection(\cite{2000mrp..book.....B}, \cite{2000mare.book.....P}) has shown that scales intermediate
between the size of a system and resistive scales can be important.  Nevertheless,
in many astrophysical settings, simple models for reconnection give time scales that are
very long, and reconnection is observed or inferred to occur on much
shorter time scales, e.g. for solar flares, more than $10^{10}$ times faster than
the theory\cite{1991JGR....96.9399D}.  This has led to the suggestion that magnetic
reconnection in the limit of vanishing resistivity might also go to a weak (discontinuous) solution,
occuring at a finite speed which is insensitive to the value of the resistivity.

The problem is best illustrated by the Sweet-Parker configuration
(\cite{1958IAUS....6..123S}, \cite{1957JGR....62..509P}), where opposing
magnetic fields interact in a thin current sheet, the reconnection layer.
This unmagnetized layer becomes a barrier to further reconnection.  In a
finite reconnection region, fluid can escape the reconnection region at
alfvenic speeds.  Because the reconnection region is thin, the reconnection
speed is reduced from the alfven speed by a factor of the ratio of the current sheet width to
the transverse system size.  In the Sweet-Park model this factor is the inverse of the square root of
the Lundquist number ($V_A L/\eta$).
The predicted sheet widths are typically extremely thin.

Petschek proposed a fast magnetic reconnection
solution (\cite{1964NASSP..50..425P}) based on the idea that  magnetic reconnection
happens in a much smaller diffusive region, called the X-point, instead
of a thin sheet.  The global structure is determined by the log of
the Lundquist number, and stationary shocks allow the fluid to convert
magnetic energy to entropy.

However, Biskamp's  simulations (\cite{1986mrt..conf...19B}) showed
that Petschek's solution is unstable when Ohmic resistivity becomes very
small.  In their two dimensional incompressible resistive MHD simulations,
they injected and ejected plasma and magnetic flux across the boundary.
They also changed the boundary condition during the simulation to
eliminate the boundary current layer.  However, considering the current sheet
formed in their simulation, the computation domain may not be big enough.  After reproducing
different scaling simulations results(\cite{1986mrt..conf...19B},
\cite{1986JGR....91.6807L}), Priest and Forbes \cite{1992JGR....9716757P}
pointed out that it is the boundary conditions that determine what
happens (including Biskamp's unstable Petscheck's simulation) and that
sufficiently free boundary conditions can make fast reconnection happen.
However, there is no self-consistent simulation of fast reconnection reported, except with
artificially enhanced local resistivity\cite{1989JGR....94.8805S}.

To reconcile the observed fast reconnection with its absence in simulations
leads to two possible resolutions: 1) ideal MHD are not the correct
equations, and long range collisionless effects are required, or 2)
assumptions about the reconnection regions are too restrictive.  This
includes the 2-dimensionality and the boundary conditions.

In exploring of the first possibility, it was found that when
integrating with the Hall term in the MHD equations, or using a kinetic
description(\cite{2001JGR...106.3715B}), it was possible to find
fast reconnection.  However, this still didn't offer any help to the
collisional system, which still has fast magnetic reconnection no
matter whether Hall term is present or not; and also the increase
of local resistivity is not generic in astrophysical environments,
which mostly has highly conducting fluids.  

For the second possibility, we note that  Lazarian \& Vishniac (LV99)
\cite{1999ApJ...517..700L} proposed a model of fast magnetic
reconnection with low amplitude turbulence.  Subsequent   simulation results
\cite{2009arXiv0903.2052K}  support  this model.  They found
that the reconnection rate depends on the amplitude of the fluctuations and
the injection scale, and that Ohmic resistivity and anomalous resistivity
do not affect the reconnection rate.  The result  that only 
the characteristics of turbulence determine the reconnection speed provides a good fit
for reconnection in  astrophysical systems.

LV99 offered a solution to fast magnetic reconnection in collisional
systems with turbulence.  In this paper, we consider a different problem,
whether we could still have fast reconnection without turbulence.
We present an example of fast magnetic reconnection in ideal three
dimensional MHD simulation in the absence of turbulence.
Here we explore a different aspect: 3-D effects and boundary
conditions.  Traditionally, simulations have searched for stationary
2-D solutions, or scaling solutions.  In the case of fast reconnection,
the geometry changes on an alfvenic time, so these assumptions might not
be applicable.  Specifically, we bypass the choice of boundary condition
by using a periodic box.

The primary constructive fast reconnection solution, the Petscheck
solution, has some peculiar aspects.  The global geometry of the flow, and
the reconnection speed, depend on the details of a microscopic X-point.
This X-point actually interacts infinitesimal matter and energy,
so it seems rather surprising that this tiny volume could affect the
global flow.  Instead, one might worry about the global flow of the system,
which dominates the energy.  We will see that this is particularly important in 
our simulations.

\section{Simulation setup}
\subsection{Physical setup}

The purpose of the simulation is to study magnetic reconnection and
its dynamics.  We start by dividing the volume in two, with each subvolume containing a uniform
magnetic field.
In a periodic volume, this results in two current sheets where reconnection
can occur.  An initial perturbation is added to trigger the
reconnection.

\subsection{Computational implementation}

Our simulations were performed on the Canadian Institute for Theoretical
Astrophysics Sunnyvale cluster: 200 Dell PE1950 compute nodes; each node
contains 2 quad core Intel(R) Xeon(R) E5310 @ 1.60GHz processors, 4GB of
RAM, and 2 gigE network interfaces.  The code \cite{2003ApJS..149..447P}
is a second-order accurate (in space and time) high-resolution total
variation diminishing (TVD) MHD parallel code.  Kinetic, thermal, and
magnetic energy are conserved and the divergence of the magnetic field
was kept zero by flux constrained transport.  There is no explicit
magnetic and viscous dissipation in the code.
The TVD constraints result in non-linear viscosity and resistivity
on the grid scale.

\subsection{Numerical setup}

\begin{figure}
\centering
\includegraphics[scale=0.3]{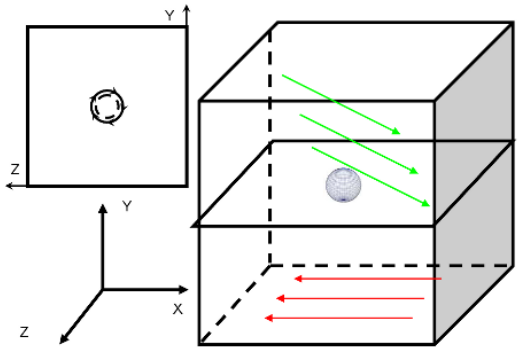}
\caption{numerical setup: the sphere in the center of the box represent the area of the rotational perturbation.
up-left is the rotational perturbation looked from YZ plane.}
\label{figure:setup}
\end{figure}

We have a reference setup, and vary numerical parameters relative to
that.  Initially the upper and lower halves of the simulation volume
are filled with uniform magnetic fields whose directions differ by 135
degrees (Figure \ref{figure:setup}).  The magnitude of the magnetic
field is the same for every cell, and $\beta$, the ratio of gas
pressure to magnetic pressure, is set to one.

There is a rotational perturbation on the interface of the magnetic
field, at the center of the box, inside a sphere of radius  0.05,
relative to the box size.  The rotational axis is nearly along the X
axis, with a small deviation, which is used to break any residual
symmetry.  We use constant specific angular momentum at the equator,
with solid body rotation on shells, which comes from the same initial
condition generator as \cite{2003ApJ...596L.207P}.
The rotational speed is set to equal to the sound speed at a radius of 0.02,
and 0.4 sound speed at the sphere's equatorial surface

We also tried adding a localized magnetic field perturbation: a random Gaussian
magnetic field, with ($\beta=1$) and correlation length is half of the box,
was added in the same region as the rotational perturbation.
Since the only dissipation is numerical, 
on the grid scale, a translational velocity \cite{2004NewA....9..443T}
was added to the simulation to increase the numerical diffusion for
all the cells  
in the box.  The reference value of the translational velocity is
equal to the sound speed 
and we measure the time (unit in CT) by box size divided by the
initial sound speed.  Varying this by a factor of 2 up or down does
not change the results.
At the beginning the Alfven speed is the same as the sound speed.  Different resolutions were tested, from $50^3$ cells to
$800^3$ cells.

\begin{figure}
\centering
\includegraphics[scale=0.4]{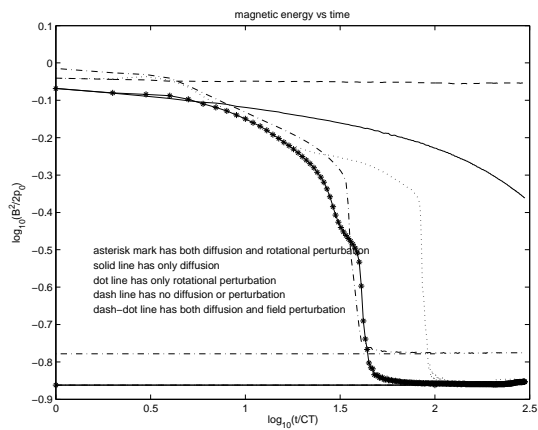}
\caption{fast reconnection for different initial conditions}
\label{figure:why_FRC}
\end{figure}

\begin{figure}
\centering
\includegraphics[scale=0.4]{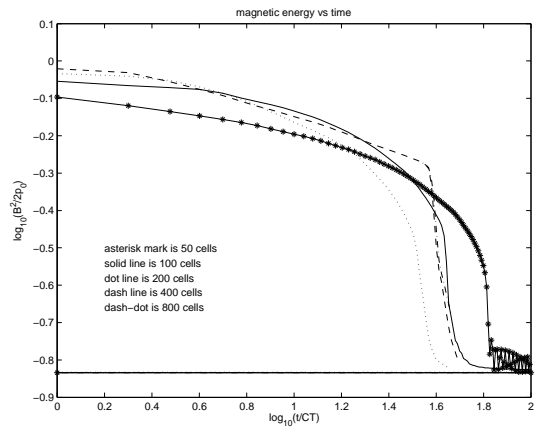}
\caption{fast reconnection for different resolutions}
\label{figure:reso_FRC}
\end{figure}

\begin{figure}
\centering
\includegraphics[scale=0.45]{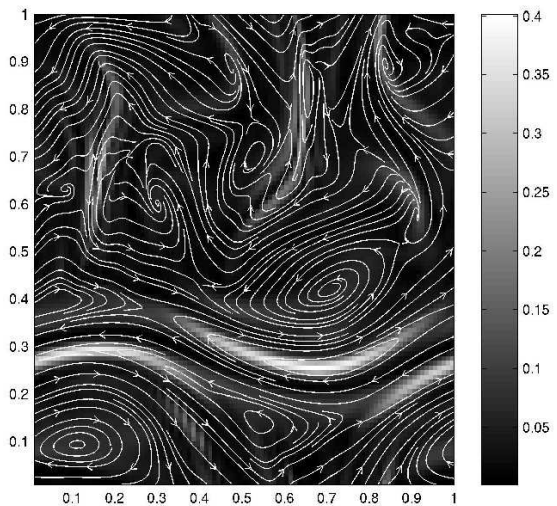}
\caption{2D snapshot during reconnection. current as background color}
\label{figure:100cells_rough_FRC}
\end{figure}

\section{Simulation results}
\subsection{Global fast magnetic reconnection}

We use the total magnetic energy as a global diagnostic
of the system.  Figure \ref{figure:why_FRC} shows  the evolution of the magnetic energy.  
The generic feature is the sudden drop of
magnetic energy, which occurs on an alfvenic box crossing time, during which
much of the magnetic energy is dissipated.  The onset of this event
depends on numerical parameters.  Due to symmetries in the code,
an absence of any initial perturbations would maintain the initial
conditions indefinitely.

We can see that when there is no forced
diffusion and no initial perturbation, the magnetic energy is almost stationary.  
When diffusion is added,
the magnetic energy decays gradually throughout the simulation.

When explicit velocity perturbations are present, all the simulations show
a sudden decrease of magnetic energy, which indicates fast magnetic
reconnection.  The common property is that they all have some initial
perturbation, either rotational or a strong localized field perturbation; and the
background diffusion only affects how early reconnection happens.
In order to make sure this fast reconnection is not related to resolution,
we simulate different resolutions, from a $50^3$ box, to a $800^3$ box,
in Figure \ref{figure:reso_FRC}.  All show fast reconnection and
the resolution only affects the time elapsed before  fast reconnection happens,
though the details of how the delay depends on resolution are still unclear.
Figure \ref{figure:100cells_rough_FRC} shows a rough two dimensional
snapshot of current($\propto\bigtriangledown\times{B}$) during fast
reconnection, with color representing the current magnitude.  It is clear that
there are some regions that have reconnection(i.e. high current value)
and we will use higher resolution to analyze them later.  So, how fast is
reconnection here? Since the magnetic energy is $\sim$ 50\% at the onset of fast
reconnection, the Alfven time is also $\sim$  CT and in the simulation,
we find that nearly 50\% of magnetic energy was released 
in one Alfven time during magnetic reconnection.  This is clearly fast reconnection by any 
reasonable criteria.

\subsection{What happens on the current sheet?}

We can see there are some regions that have large currents, and the
reconnection should happen there.  Now we use high resolution (e.g. 800
cells) to investigate what exactly happens there.  We want to show a
snapshot close to the current sheet to see how flow evolves and what the
magnetic field geometry looks like near the current sheet. We subtract
the average value for both magnetic field and velocity in the region close
to the current sheet.  This places us in the frame comoving with the fluid.
The mean magnetic field does not participate in the dynamics of reconnection,
so its removal allows us to see the dynamics more clearly.

\begin{sidewaysfigure}
\centering
\includegraphics[scale=1.2]{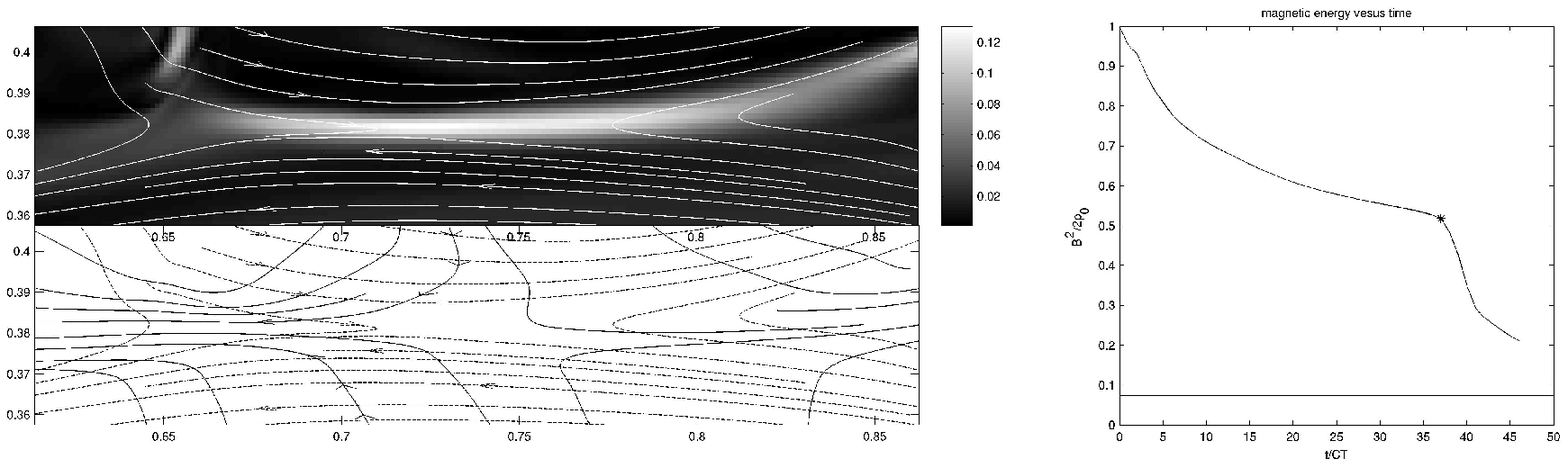}
\caption{snapshot of magnetic field line on the background of current, and snapshot of both magnetic and velocity field line, and $B^2$ at 37 CT}
\label{figure:and1_800_185}
\end{sidewaysfigure}

\begin{sidewaysfigure}
\centering
\includegraphics[scale=1.2]{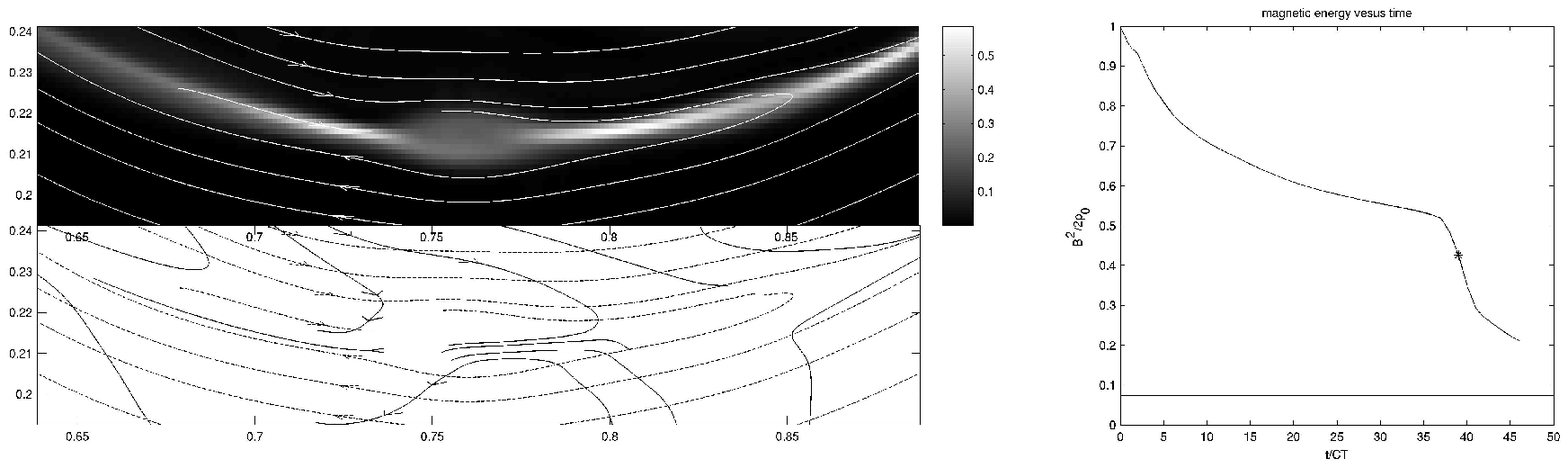}
\caption{snapshot of magnetic field line on the background of current, and snapshot of both magnetic and velocity field line, and $B^2$ at 39 CT}
\label{figure:and1_800_195}
\end{sidewaysfigure}

\begin{sidewaysfigure}
\centering
\includegraphics[scale=1.2]{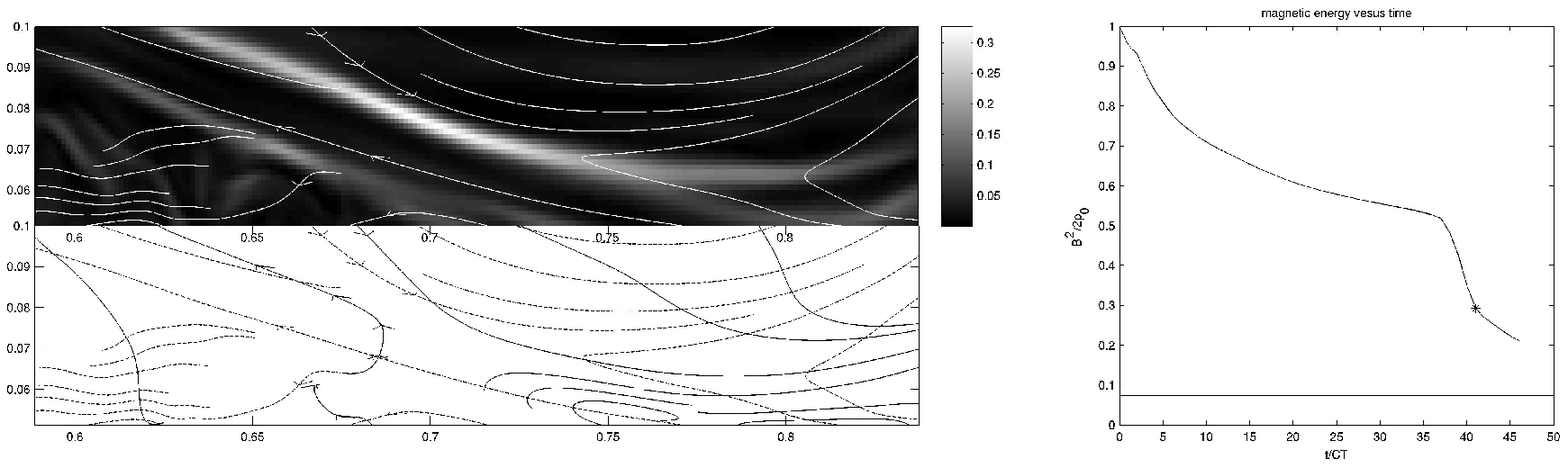}
\caption{snapshot of magnetic field line on the background of current, and snapshot of both magnetic and velocity field line, and $B^2$ at 41 CT}
\label{figure:and1_800_205}
\end{sidewaysfigure}

We present snapshots of three different times during the reconnection:
one at the beginning, one at the middle and one at the end.  Each time
step snapshot contains three graphs, with the upper left one has current
magnitude as background color and white line represents magnetic field line,
and the lower left one is a snapshot of both magnetic field (blue dash) and
velocity field (red solid), and the right one is the corresponding magnetic
energy plot.  Figure \ref{figure:and1_800_185} is the beginning; Figure
\ref{figure:and1_800_195} is the middle; Figure \ref{figure:and1_800_205}
is the end;

\begin{figure}
\centering
\includegraphics[scale=0.45]{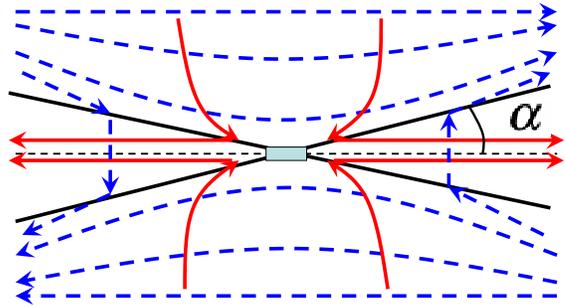}
\caption{geometry of Petscheck solution}
\label{figure:petscheck}
\end{figure}

It is easy to find that the snapshot of both magnetic field line and
velocity field line in figure \ref{figure:and1_800_185} looks like Figure
\ref{figure:petscheck} \cite{1964NASSP..50..425P}, which is the geometry
of Petschek's solution for fast magnetic reconnection.  The X-point, which
is the reconnection region, is small and at the center.  The tangent of
the angle $\alpha$ represent the ratio of inflow to outflow.

\subsection{What happens globally?}

We show the long term and global 2D evolution of both velocity field
lines and magnetic fields for the $400^3$ simulations, starting from
the beginning until reconnection completes.  These plots are analogous
to the plots in the previous section: the left one is the snapshot of
both magnetic and velocity field lines; the center one is the snapshot
of magnetic field lines with current as the background color; and the
corresponding magnetic energy is also included on the right.  At the
beginning, the magnetic field lines are opposite and there is no velocity
field.  Then the initial rotational perturbation induces two reconnected
regions with closed magnetic field loops, one at each interface.
The closed loops are fed by a slow X-point at each interface.
Noting that there is a mean field perpendicular to the plotted surface,
these loops are actually twists in the perpendicular magnetic field.
In the bulk region between the interfaces, the parallel magnetic fields
are not yet disturbed much by the perturbation.  

\begin{figure}
\centering
\includegraphics[scale=0.5]{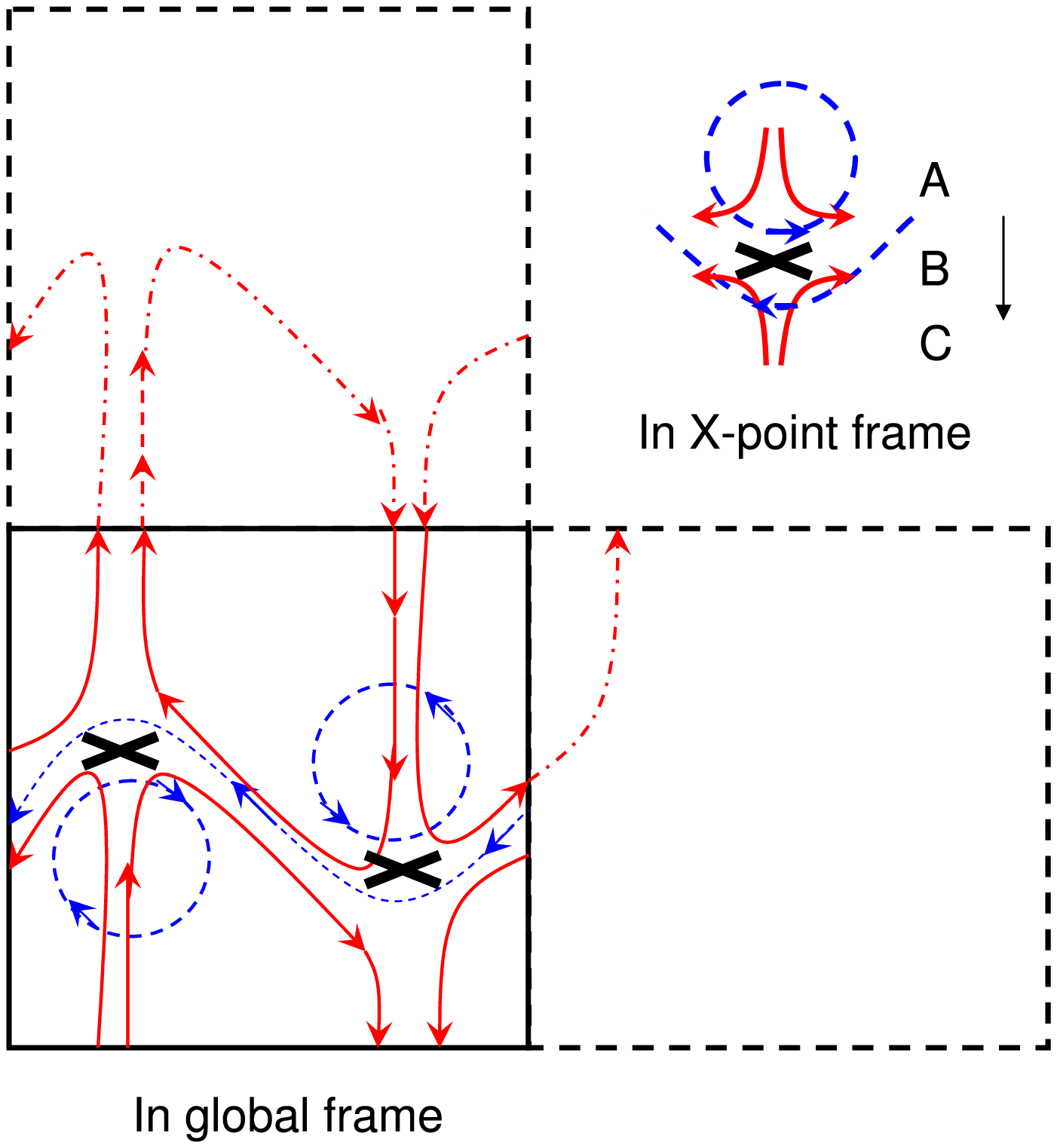}
\caption{geometry of global configuration}
\label{figure:global_c}
\end{figure}

\begin{sidewaysfigure}
\includegraphics[scale=1]{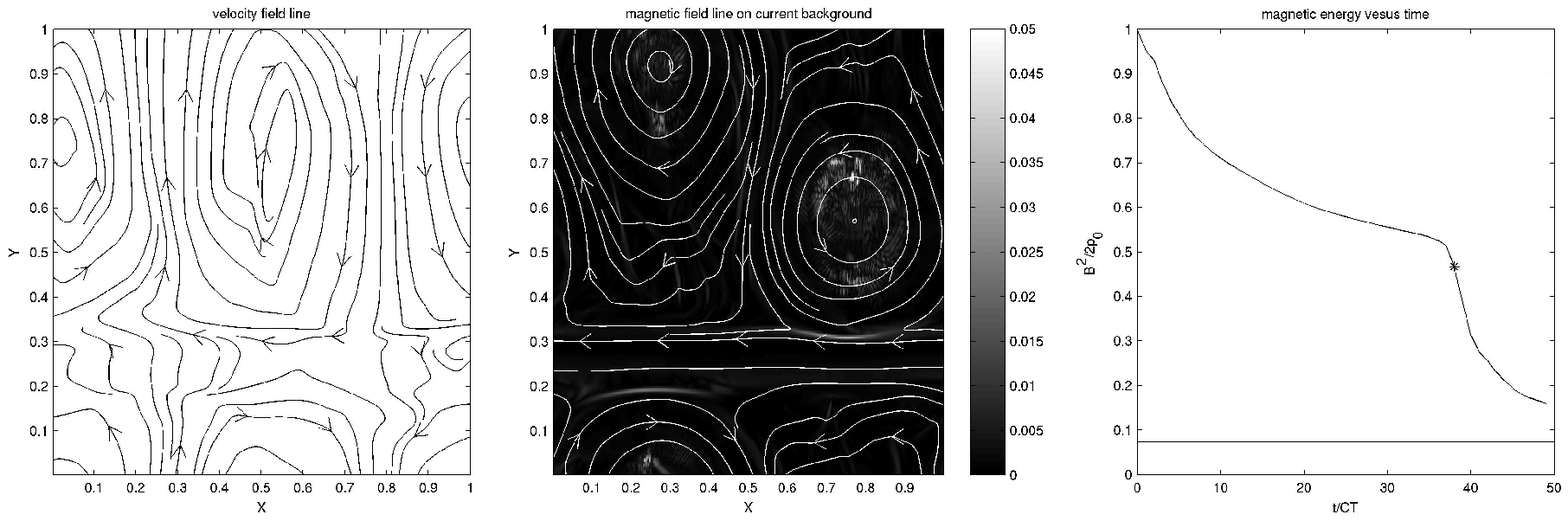}
\caption{snapshot of magnetic field line on the background of current, and snapshot of both magnetic and velocity field line, and $B^2$ at 38 CT for 400 cells}
\label{figure:and1_400_38}
\end{sidewaysfigure}

In Figure \ref{figure:and1_400_38} we can see the loops to move into the
X-point of the opposing loop, and strong interactions occur.  The fluid
forms two large circular cells, offset from the magnetic loops.  The
energy to drive the fluid flow comes from the reconnection energy of the
magnetic field.  This flow pattern enhances the reconnection by driving
the fluid through the X point.  We illustrate the fast reconnection flows
in Figure \ref{figure:global_c}. Blue dash circles with arrows represent the
magnetic loops.  The red field lines with arrows represent the velocity
field.  There are two big black X's in the global frame, which represent
the X point for reconnection.  Because we are using periodic boundary
condition, we extend the simulation box picture to two other directions, to
make the global flow easier to understand.  Red solid lines represent
the velocity field in the real box, and red dash-dot line represents
the field line in the extended box.  

Reconnection is a local process in the global flow field.  To see
that, we need to boost into the comoving frame.  Let's take the right
magnetic twist for example: In global frame, the flow on the right
all moves downwards, with the magnetic twist moving at the highest
speed.  The X-point is like a saddle point for the flow: the fluid
converges vertically, and diverges horizontally.
In the X-point frame,
setting the velocity at B to zero, A will move down and C will move up,
which supports the conditions for reconnection.

\begin{sidewaysfigure}
\includegraphics[scale=1]{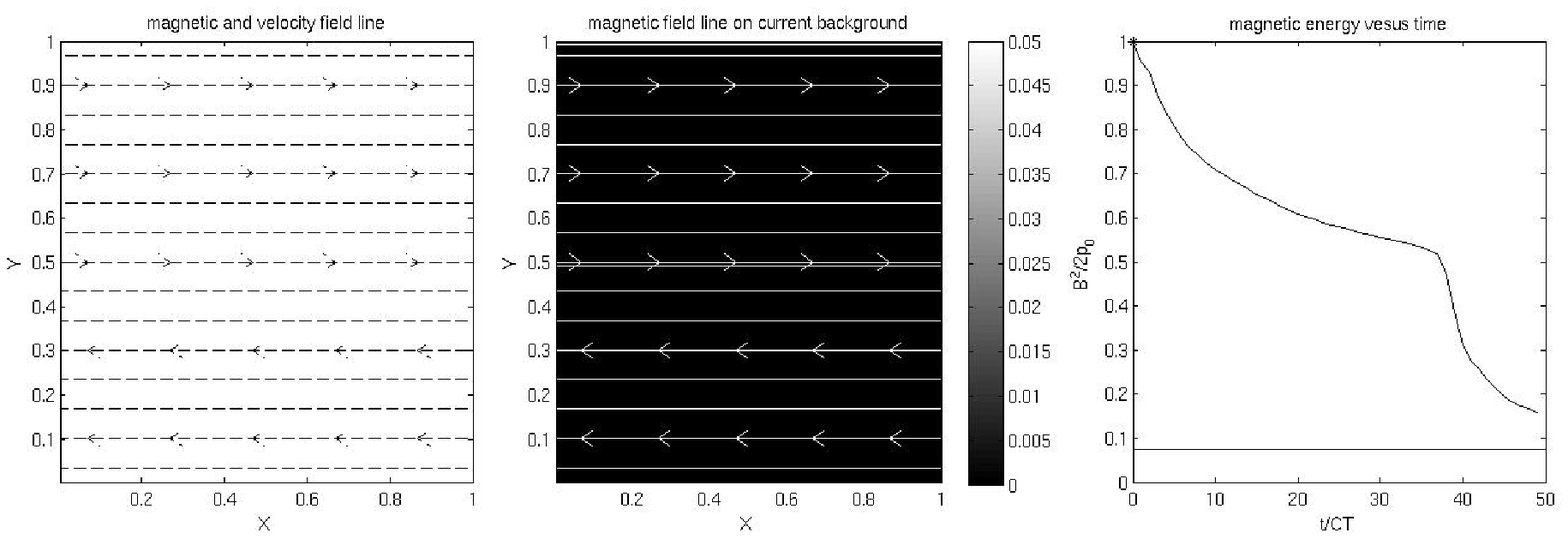}
\caption{snapshot of magnetic field line on the background of current, and snapshot of both magnetic and velocity field line, and $B^2$ at 0 CT for 400 cells}
\label{figure:and1_400_0}
\end{sidewaysfigure}

\begin{sidewaysfigure}
\includegraphics[scale=1]{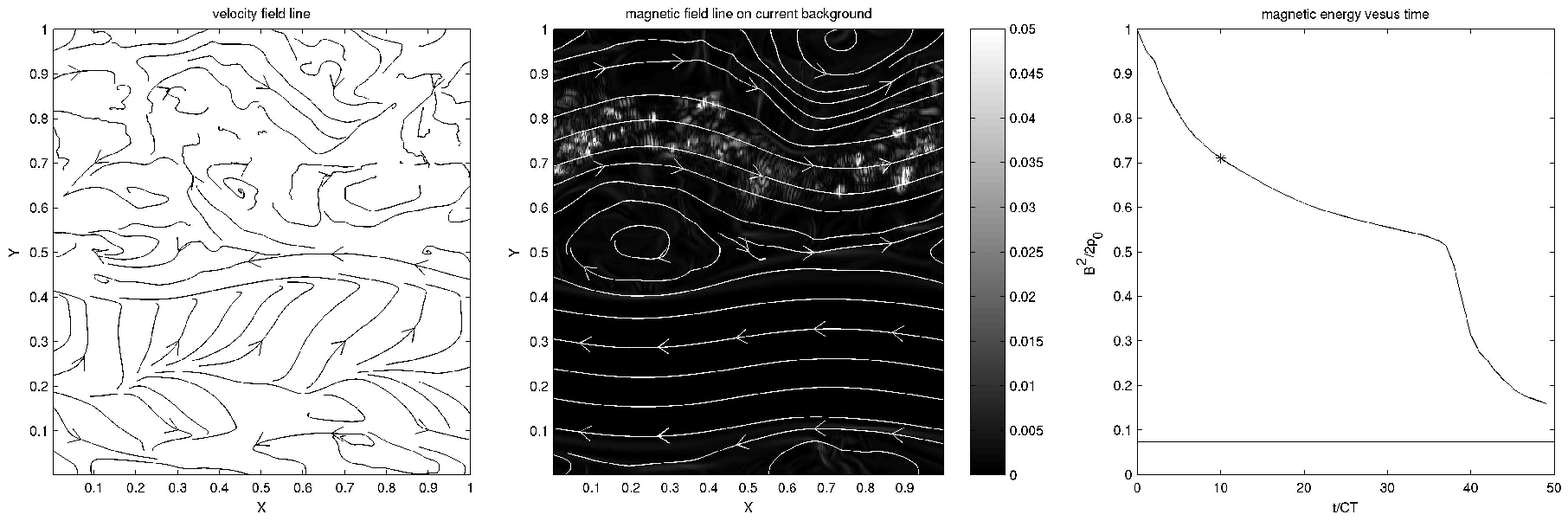}
\caption{snapshot of magnetic field line on the background of current, and snapshot of both magnetic and velocity field line, and $B^2$ at 10 CT for 400 cells}
\label{figure:and1_400_10}
\end{sidewaysfigure}

\begin{sidewaysfigure}
\includegraphics[scale=1]{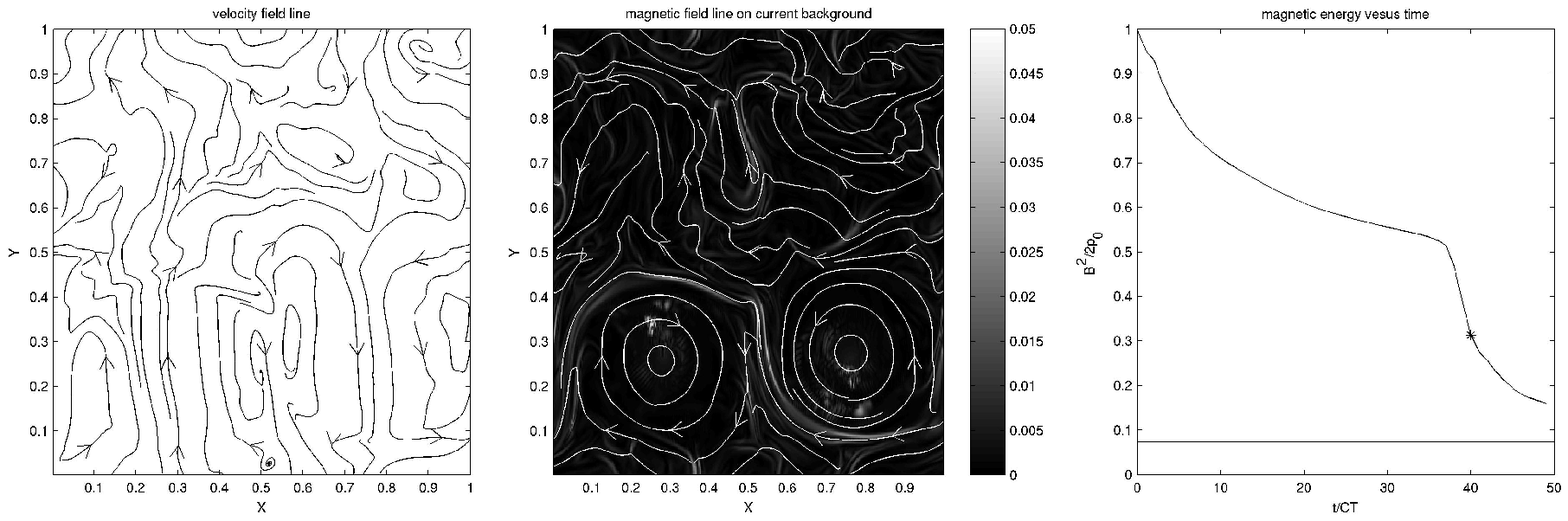}
\caption{snapshot of magnetic field line on the background of current, and snapshot of both magnetic and velocity field line, and $B^2$ at 40 CT for 400 cells}
\label{figure:and1_400_40}
\end{sidewaysfigure}

\begin{sidewaysfigure}
\includegraphics[scale=1]{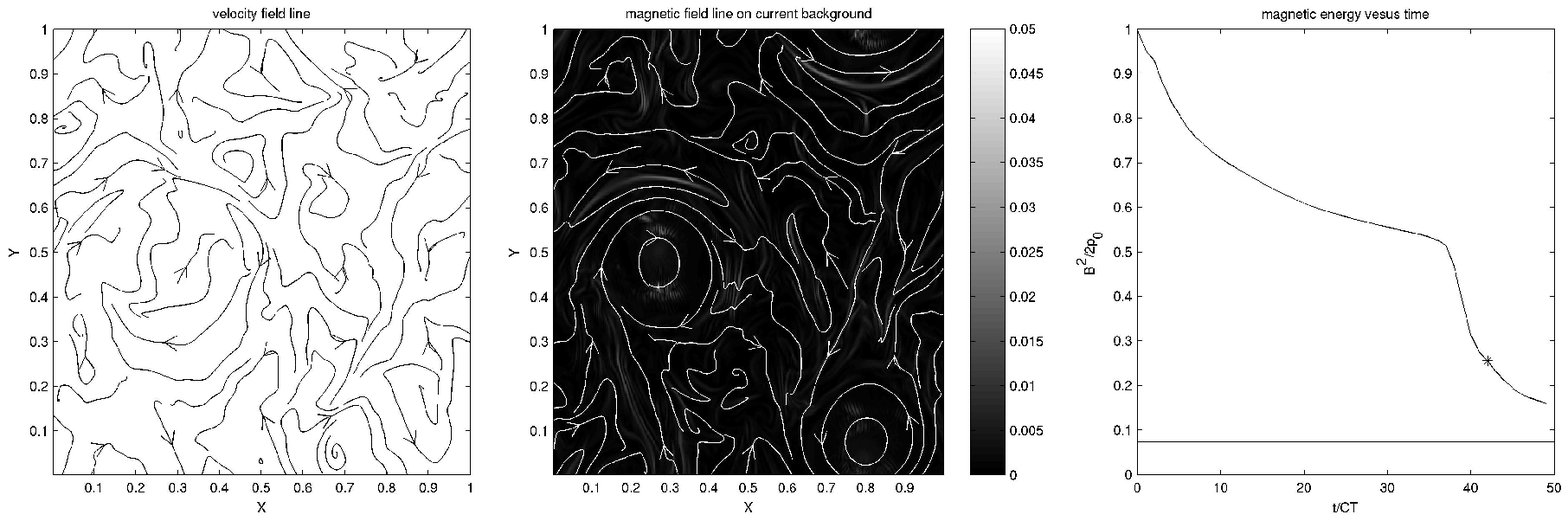}
\caption{snapshot of magnetic field line on the background of current, and snapshot of both magnetic and velocity field line, and $B^2$ at 42 CT for 400 cells}
\label{figure:and1_400_42}
\end{sidewaysfigure}

\section{Discussion}

To summarize, we have found a global flow pattern which reinforces
 X-point reconnection, and the resulting fast reconnection in turn drives the
global flow pattern.  The basic picture is two dimensional.  We did
find that a pure 2-D simulation does not show this fast reconnection.
This is easy to understand, since the reconnected field loops are loaded
with matter, and would require resistivity to dissipate.  In 3-D, these
loops are twists which are unstable to a range of instabilities, allowing
the field loops to collapse.  So three basic ingredients are
needed:
1. A global flow which keeps the field lines outside the X-point
at a large opening angle to allow the reconnected fluid to escape, and
avoid the Sweet-Parker time scale.
2. The reconnection energy drives this global flow
3. A three dimensional instability allows closed (reconnected) field
lines to collapse, releasing all the energy stored in the field.

The problem  described here has two geometric dimensionless
parameters: the 2 axis ratios of the periodic box.  In addition,
there are a number of numerical parameters.  We have varied them
to study their effects.

Extending the box in the Y direction (separation between reconnection
regions) shuts off this instability, which might be expected: there
are no global flows possible if the two interaction regions are too
far separated.  We found the threshold to be $Y < 1.2Z$.  In the other
direction, there appears to be no limit to make $Y << Z$.  Increasing
the size of the Z dimension does not diminish this instability. There is also a
dependence on X (extend along field symmetry axis).  Shortening it to one
grid cell protects the topology of field loops, and reconnection
is not observed in 2-D simulations.


We changed different initial condition to see whether the fast
reconnection is sensitive to how the initial setup is.  After changing
the angle of the opposite magnetic field(from beyond 90 degree to 180
degree), the strength of the rotational perturbation, and axis of the
rotational perturbation, we found
that the fast reconnection still appeared.  The boundary condition is
kept periodic and we found that the evolution of fluid dynamics
of different initial conditions are similar.


It can be seen that the fast reconnection happens at the two interfaces of
the straight magnetic field at the same time, with a magnetic twist moving
towards it on each side.  They are not head-on collision on the magnetic
field, but a little separated in transverse direction.  This special
geometry helps the magnetic reconnection happen fast, since each
magnetic twist pushes the field line, it also affect the velocity
field at the other side and it helps to increase the outflow speed.
If we look back to Sweet-Parker's solution(\cite{1958IAUS....6..123S},
\cite{1957JGR....62..509P}), the main problem is that the current
sheet is so thin, that even if one accelerates the outflow to Alfven
speed, the mass of outflow is still small, which slows down the speed
of the reconnection.  Petschek's 
configuration\cite{1964NASSP..50..425P} can resolve this problem with
a small  reconnection region and finite opening angle for the outflow.
In
our simulation  the speed of the outflow is further increased by the
feedback between the two reconnection regions.


The solar flare reconnection time scale is about Alfven time
scale\cite{1991JGR....96.9399D}, which is the order of seconds to minutes.

If there is only magnetic diffusivity($\eta$) present, the diffusive
time is $\tau_{D}=L^2/\eta$, with $L$ is the characteristic length.
Taking the values from \cite{1991JGR....96.9399D}, $L=1000km$ and $\eta$
is $10^{-3}{m^2}s^{-1}$, $\tau_{D}$ is $10^{15}s$.

Sweet-Parker's thin current sheet proposed a reconnection
time as $\tau_{SP}=L/(V_{Ai}/R_{mi}^{1/2})$, with
$R_{mi}=L\upsilon_{Ai}/\eta$. This makes the reconnection time about
$10^5$ Alfven times.

Petschek's configuration has a reconnection time as
$\tau_{P}=L/(\alpha\upsilon_{A})$, with $\alpha$ is between 0.01 and 0.1
and Alfven speed $\sim 100km/s$, and this makes the time scale as $100-1000s$.

Our fast reconnection time has the order of Alfven time scale, and Alfven
time $\tau_{A}=L/\upsilon_{A}$, which is the same order as observed time
scales of $20-60s$ \cite{1991JGR....96.9399D}.  Furthermore, comparing
to LV99, no turbulence is needed or added in our simulations.
Our fast magnetic reconnection time scale is qualitatively similar to
the energy release time scale for solar flares.

\section{Summary}

We present evidence for fast magnetic reconnection in a global three dimensional ideal
magnetohydrodynamics simulation without any sustained external driving.
These global simulations are self-contained, and do not rely on specified boundary
conditions.  We have quantified ranges in parameter
space where fast reconnection is generic.  The reconnection is
Petscheck-like, and fast, meaning that nearly half of the magnetic energy is
released in one Alfven time.

This example of  fast reconnection example relies on two interacting reconnection
regions in a periodic box.   It is an intrinsically three dimensional effect.  Our
interpretation is that the Petschek-like X-point angles are not determined
by microscopic properties at an infinitesimal boundary where no energy
is present, but rather by the global flow far away from the X-point.  Whether or not such configurations 
are natural in an open system remains to be seen.

\begin{acknowledgments}
We would like to thank Christopher D. Matzner for helpful comments.  
The computations were performed on CITA's Sunnyvale clusters which are funded by the Canada Foundation for Innovation, the Ontario Innovation Trust, and the Ontario Research Fund.
The work of ETV and UP is supported by the National Science and Engineering Research 
Council of Canada.
\end{acknowledgments}

\bibliography{bpangbib}

\begin{thebibliography}{16}
\expandafter\ifx\csname natexlab\endcsname\relax\def\natexlab#1{#1}\fi
\expandafter\ifx\csname bibnamefont\endcsname\relax
  \def\bibnamefont#1{#1}\fi
\expandafter\ifx\csname bibfnamefont\endcsname\relax
  \def\bibfnamefont#1{#1}\fi
\expandafter\ifx\csname citenamefont\endcsname\relax
  \def\citenamefont#1{#1}\fi
\expandafter\ifx\csname url\endcsname\relax
  \def\url#1{\texttt{#1}}\fi
\expandafter\ifx\csname urlprefix\endcsname\relax\def\urlprefix{URL }\fi
\providecommand{\bibinfo}[2]{#2}
\providecommand{\eprint}[2][]{\url{#2}}

\bibitem[{\citenamefont{{Biskamp}}(2000)}]{2000mrp..book.....B}
\bibinfo{author}{\bibfnamefont{D.}~\bibnamefont{{Biskamp}}},
  \emph{\bibinfo{title}{{Magnetic Reconnection in Plasmas}}}
  (\bibinfo{year}{2000}).

\bibitem[{\citenamefont{{Priest} and {Forbes}}(2000)}]{2000mare.book.....P}
\bibinfo{author}{\bibfnamefont{E.}~\bibnamefont{{Priest}}} \bibnamefont{and}
  \bibinfo{author}{\bibfnamefont{T.}~\bibnamefont{{Forbes}}},
  \emph{\bibinfo{title}{{Magnetic Reconnection}}} (\bibinfo{year}{2000}).

\bibitem[{\citenamefont{{Dere} et~al.}(1991)\citenamefont{{Dere}, {Bartoe},
  {Brueckner}, {Ewing}, and {Lund}}}]{1991JGR....96.9399D}
\bibinfo{author}{\bibfnamefont{K.~P.} \bibnamefont{{Dere}}},
  \bibinfo{author}{\bibfnamefont{J.-D.~F.} \bibnamefont{{Bartoe}}},
  \bibinfo{author}{\bibfnamefont{G.~E.} \bibnamefont{{Brueckner}}},
  \bibinfo{author}{\bibfnamefont{J.}~\bibnamefont{{Ewing}}}, \bibnamefont{and}
  \bibinfo{author}{\bibfnamefont{P.}~\bibnamefont{{Lund}}},
  \bibinfo{journal}{jgr} \textbf{\bibinfo{volume}{96}}, \bibinfo{pages}{9399}
  (\bibinfo{year}{1991}).

\bibitem[{\citenamefont{{Sweet}}(1958)}]{1958IAUS....6..123S}
\bibinfo{author}{\bibfnamefont{P.~A.} \bibnamefont{{Sweet}}}, in
  \emph{\bibinfo{booktitle}{Electromagnetic Phenomena in Cosmical Physics}},
  edited by \bibinfo{editor}{\bibfnamefont{B.}~\bibnamefont{{Lehnert}}}
  (\bibinfo{year}{1958}), vol.~\bibinfo{volume}{6} of
  \emph{\bibinfo{series}{IAU Symposium}}, pp. \bibinfo{pages}{123--+}.

\bibitem[{\citenamefont{{Parker}}(1957)}]{1957JGR....62..509P}
\bibinfo{author}{\bibfnamefont{E.~N.} \bibnamefont{{Parker}}},
  \bibinfo{journal}{jgr} \textbf{\bibinfo{volume}{62}}, \bibinfo{pages}{509}
  (\bibinfo{year}{1957}).

\bibitem[{\citenamefont{{Petschek}}(1964)}]{1964NASSP..50..425P}
\bibinfo{author}{\bibfnamefont{H.~E.} \bibnamefont{{Petschek}}},
  \bibinfo{journal}{NASA Special Publication} \textbf{\bibinfo{volume}{50}},
  \bibinfo{pages}{425} (\bibinfo{year}{1964}).

\bibitem[{\citenamefont{{Biskamp}}(1986)}]{1986mrt..conf...19B}
\bibinfo{author}{\bibfnamefont{D.}~\bibnamefont{{Biskamp}}}, in
  \emph{\bibinfo{booktitle}{Magnetic Reconnection and Turbulence}}, edited by
  \bibinfo{editor}{\bibfnamefont{M.~A.} \bibnamefont{{Dubois}}},
  \bibinfo{editor}{\bibfnamefont{D.}~\bibnamefont{{Gr{\'e}sellon}}},
  \bibnamefont{and} \bibinfo{editor}{\bibfnamefont{M.~N.}
  \bibnamefont{{Bussac}}} (\bibinfo{year}{1986}), pp. \bibinfo{pages}{19--+}.

\bibitem[{\citenamefont{{Lee} and {Fu}}(1986)}]{1986JGR....91.6807L}
\bibinfo{author}{\bibfnamefont{L.~C.} \bibnamefont{{Lee}}} \bibnamefont{and}
  \bibinfo{author}{\bibfnamefont{Z.~F.} \bibnamefont{{Fu}}},
  \bibinfo{journal}{jgr} \textbf{\bibinfo{volume}{91}}, \bibinfo{pages}{6807}
  (\bibinfo{year}{1986}).

\bibitem[{\citenamefont{{Priest} and {Forbes}}(1992)}]{1992JGR....9716757P}
\bibinfo{author}{\bibfnamefont{E.~R.} \bibnamefont{{Priest}}} \bibnamefont{and}
  \bibinfo{author}{\bibfnamefont{T.~G.} \bibnamefont{{Forbes}}},
  \bibinfo{journal}{jgr} \textbf{\bibinfo{volume}{97}}, \bibinfo{pages}{16757}
  (\bibinfo{year}{1992}).

\bibitem[{\citenamefont{{Scholer}}(1989)}]{1989JGR....94.8805S}
\bibinfo{author}{\bibfnamefont{M.}~\bibnamefont{{Scholer}}},
  \bibinfo{journal}{jgr} \textbf{\bibinfo{volume}{94}}, \bibinfo{pages}{8805}
  (\bibinfo{year}{1989}).

\bibitem[{\citenamefont{{Birn} et~al.}(2001)\citenamefont{{Birn}, {Drake},
  {Shay}, {Rogers}, {Denton}, {Hesse}, {Kuznetsova}, {Ma}, {Bhattacharjee},
  {Otto} et~al.}}]{2001JGR...106.3715B}
\bibinfo{author}{\bibfnamefont{J.}~\bibnamefont{{Birn}}},
  \bibinfo{author}{\bibfnamefont{J.~F.} \bibnamefont{{Drake}}},
  \bibinfo{author}{\bibfnamefont{M.~A.} \bibnamefont{{Shay}}},
  \bibinfo{author}{\bibfnamefont{B.~N.} \bibnamefont{{Rogers}}},
  \bibinfo{author}{\bibfnamefont{R.~E.} \bibnamefont{{Denton}}},
  \bibinfo{author}{\bibfnamefont{M.}~\bibnamefont{{Hesse}}},
  \bibinfo{author}{\bibfnamefont{M.}~\bibnamefont{{Kuznetsova}}},
  \bibinfo{author}{\bibfnamefont{Z.~W.} \bibnamefont{{Ma}}},
  \bibinfo{author}{\bibfnamefont{A.}~\bibnamefont{{Bhattacharjee}}},
  \bibinfo{author}{\bibfnamefont{A.}~\bibnamefont{{Otto}}},
  \bibnamefont{et~al.}, \bibinfo{journal}{jgr} \textbf{\bibinfo{volume}{106}},
  \bibinfo{pages}{3715} (\bibinfo{year}{2001}).

\bibitem[{\citenamefont{{Lazarian} and {Vishniac}}(1999)}]{1999ApJ...517..700L}
\bibinfo{author}{\bibfnamefont{A.}~\bibnamefont{{Lazarian}}} \bibnamefont{and}
  \bibinfo{author}{\bibfnamefont{E.~T.} \bibnamefont{{Vishniac}}},
  \bibinfo{journal}{ApJ} \textbf{\bibinfo{volume}{517}}, \bibinfo{pages}{700}
  (\bibinfo{year}{1999}), \eprint{arXiv:astro-ph/9811037}.

\bibitem[{\citenamefont{{Kowal} et~al.}(2009)\citenamefont{{Kowal}, {Lazarian},
  {Vishniac}, and {Otmianowska-Mazur}}}]{2009arXiv0903.2052K}
\bibinfo{author}{\bibfnamefont{G.}~\bibnamefont{{Kowal}}},
  \bibinfo{author}{\bibfnamefont{A.}~\bibnamefont{{Lazarian}}},
  \bibinfo{author}{\bibfnamefont{E.~T.} \bibnamefont{{Vishniac}}},
  \bibnamefont{and}
  \bibinfo{author}{\bibfnamefont{K.}~\bibnamefont{{Otmianowska-Mazur}}},
  \bibinfo{journal}{ArXiv e-prints}  (\bibinfo{year}{2009}),
  \eprint{0903.2052}.

\bibitem[{\citenamefont{{Pen} et~al.}(2003{\natexlab{a}})\citenamefont{{Pen},
  {Arras}, and {Wong}}}]{2003ApJS..149..447P}
\bibinfo{author}{\bibfnamefont{U.-L.} \bibnamefont{{Pen}}},
  \bibinfo{author}{\bibfnamefont{P.}~\bibnamefont{{Arras}}}, \bibnamefont{and}
  \bibinfo{author}{\bibfnamefont{S.}~\bibnamefont{{Wong}}},
  \bibinfo{journal}{ApJs} \textbf{\bibinfo{volume}{149}}, \bibinfo{pages}{447}
  (\bibinfo{year}{2003}{\natexlab{a}}).

\bibitem[{\citenamefont{{Pen} et~al.}(2003{\natexlab{b}})\citenamefont{{Pen},
  {Matzner}, and {Wong}}}]{2003ApJ...596L.207P}
\bibinfo{author}{\bibfnamefont{U.}~\bibnamefont{{Pen}}},
  \bibinfo{author}{\bibfnamefont{C.~D.} \bibnamefont{{Matzner}}},
  \bibnamefont{and} \bibinfo{author}{\bibfnamefont{S.}~\bibnamefont{{Wong}}},
  \bibinfo{journal}{ApJ} \textbf{\bibinfo{volume}{596}}, \bibinfo{pages}{L207}
  (\bibinfo{year}{2003}{\natexlab{b}}), \eprint{arXiv:astro-ph/0304227}.

\bibitem[{\citenamefont{{Trac} and {Pen}}(2004)}]{2004NewA....9..443T}
\bibinfo{author}{\bibfnamefont{H.}~\bibnamefont{{Trac}}} \bibnamefont{and}
  \bibinfo{author}{\bibfnamefont{U.-L.} \bibnamefont{{Pen}}},
  \bibinfo{journal}{New Astronomy} \textbf{\bibinfo{volume}{9}},
  \bibinfo{pages}{443} (\bibinfo{year}{2004}), \eprint{arXiv:astro-ph/0309599}.

\end{thebibliography}

\end{document}